\documentclass[journal]{IEEEtran}
\pdfoutput = 1
\usepackage{amssymb}
\usepackage{amsthm}
\usepackage{bm}
\usepackage{cite}
\usepackage{amsmath}
\usepackage{array}
\usepackage{booktabs}
\usepackage{graphicx}
\usepackage{float}
\usepackage{amsfonts}
\usepackage{subfigure}
\usepackage {color}
\usepackage{stfloats}
\usepackage{hyperref}
\usepackage[ruled,linesnumbered]{algorithm2e}  

\begin{document}

\bibliographystyle{IEEEtran}

\title{\huge Reconfigurable Massive MIMO: Harnessing the Power of the Electromagnetic Domain for Enhanced Information Transfer}

\author{Keke Ying, Zhen Gao,~\IEEEmembership{Member,~IEEE}, Sheng Chen,~\IEEEmembership{Fellow,~IEEE}, Xinyu Gao, \\ Michail Matthaiou,~\IEEEmembership{Fellow,~IEEE}, Rui Zhang,~\IEEEmembership{Fellow,~IEEE}, and Robert Schober,~\IEEEmembership{Fellow,~IEEE}
\thanks{K. Ying is with the School of Information and Electronics, Beijing Institute of Technology, Beijing 100081, China (E-mail: ykk@bit.edu.cn).} %
\thanks{Z. Gao is with the Advanced Research Institute of Multidisciplinary Science and School of Information and Electronics, Beijing Institute of Technology, Beijing 100081, China (E-mail: gaozhen16@bit.edu.cn).} %
\thanks{S. Chen is with the School of Electronics and Computer Science, University of Southampton, Southampton SO17 1BJ, U.K. (E-mail: sqc@ecs.soton.ac.uk).} %
\thanks{X. Gao is with the Huawei Technologies Company Ltd., Beijing 100085, China (E-mail: gaoxinyu@huawei.com).} %
\thanks{M. Matthaiou is with the Centre for Wireless Innovation, Queen's University Belfast, Belfast BT3 9DT, U.K. (E-mail: m.matthaiou@qub.ac.uk).} %
\thanks{R. Zhang is with the School of Science and Engineering, the Chinese University of Hong Kong (Shenzhen), China, 518172, and the Shenzhen Research Institute of Big Data  (e-mail: rzhang@cuhk.edu.cn). He is also with the Department of Electrical and Computer Engineering, National University of Singapore, Singapore 117583.} %
\thanks{R. Schober is with the Institute for Digital Communications, Friedrich-Alexander-University Erlangen-Nürnberg, 91054 Erlangen, Germany (E-mail: robert.schober@fau.de).} %
\vspace*{-5mm}
}

\maketitle

\begin{abstract}
The capacity of commercial massive multiple-input multiple-output (mMIMO) systems is constrained by the limited array aperture at the base station, and cannot meet the ever-increasing traffic demands of wireless networks. Given the array aperture, holographic MIMO with infinitesimal antenna spacing can maximize the capacity, but is physically unrealizable. As a promising alternative, reconfigurable mMIMO is proposed to harness the unexploited power of the electromagnetic (EM) domain for enhanced information transfer. Specifically, the reconfigurable pixel antenna technology provides each antenna with an adjustable EM radiation (EMR) pattern, introducing extra degrees of freedom for information transfer in the EM domain. In this article, we present the concept and benefits of availing the EMR domain for mMIMO transmission. Moreover, we propose a viable architecture for reconfigurable mMIMO systems, and the associated system model and downlink precoding are also discussed. In particular, a three-level precoding scheme is proposed, and simulation results verify its considerable spectral and energy efficiency advantages compared to traditional mMIMO systems. Finally, we further discuss the challenges, insights, and prospects of deploying reconfigurable mMIMO, along with the associated hardware, algorithms, and fundamental theory.
\end{abstract}


\section{Overview of MIMO Systems}\label{Sec:Intro}

\IEEEPARstart{I}{n} 5G communication systems and beyond, multiple-input multiple-output (MIMO) technology plays an ever-increasingly important role. Expanding the array aperture and integrating denser antenna elements within limited physical space are two possible approaches for improving the throughput of MIMO systems. In the first approach, massive MIMO (mMIMO) and extra-large scale MIMO (XL-MIMO) emerge as the key technologies for exploiting the spatial domain resources \cite{XL-MIMO}. In the second approach, holographic MIMO (HMIMO) has attracted increasing interest thanks to its ability to provide unprecedented flexibility in manipulating the electromagnetic (EM) field \cite{Holograhic}. 

Although numerous studies have been conducted, mMIMO and XL-MIMO still face hardware implementation challenges. Existing MIMO architectures can be mainly categorized into fully-digital and analog/digital hybrid ones. Due to the prohibitive cost of equipping each antenna element with a dedicated radio frequency (RF) chain in a fully-digital array (FDA), various hybrid architectures have been proposed to trade off between hardware cost and system performance \cite{HBF_survey}. By introducing a phase-shift network, a fully-connected array (FCA) divides the original high-dimensional digital domain processing into a high-dimensional analog domain operation and a low-dimensional digital domain processing, so that the number of RF chains can be significantly reduced. To further reduce the number of phase shifters, the sub-connected array (SCA) has been proposed. From FDA to FCA to SCA, the hardware design becomes easier to implement at the cost of reduced degrees of freedom (DoFs) for the associated signal processing. Although various sophisticated precoding algorithms have been proposed to compensate for this performance degradation, how to achieve a satisfactory trade-off between hardware cost and achievable performance remains an open problem.

On the other hand, as a theoretical concept, HMIMO aspires to provide maximum DoFs for the manipulation of the EM field. In contrast to conventional MIMO systems employing half-wavelength critical antenna spacing, HMIMO systems are assumed to have a continuous aperture surface with infinitesimal antenna spacing, and each point on the surface has theoretically independent adjustability of the corresponding excitation current. As a result, HMIMO is expected to have the capability to flexibly generate any current density distribution on this aperture surface. Therefore, it should be able to customize any desired EM properties (e.g., polarization, radiation pattern, etc). However, it could be practically challenging to engineer such idealized HMIMO systems since the mutual coupling effect becomes more severe as the antenna spacing decreases. Moreover, it is also difficult to realize EM-level manipulations in existing MIMO systems, since the EM properties of their antennas are fixed once the antenna has been designed and fabricated.

To break the performance limits of aperture-restricted MIMO, this article proposes a feasible hardware architecture called \textit{reconfigurable mMIMO} (R-mMIMO). Different from other reconfigurable antenna systems, which directly control the amplitude/phase response of an the incident signal, R-mMIMO can actively change the EM properties of the radiating antennas, and thereby indirectly influence the transmission channel. In the following sections, we first provide a simple example to illustrate the theoretical performance gain realized with R-mMIMO by exploiting the EM radiation (EMR) domain. Then, to showcase the practical feasibility of R-mMIMO, a detailed architecture of the proposed R-mMIMO is presented, and differences compared to traditional mMIMO (T-mMIMO) architectures are highlighted. Subsequently, the EMR domain precoding problem for R-mMIMO systems is investigated, and the spectral efficiency (SE) and energy efficiency (EE) of a viable R-mMIMO structure are compared to those of T-mMIMO designs. Finally, some open research directions and challenges for R-mMIMO are discussed.

\section{EM Radiation Gain of R-mMIMO Systems}\label{S2}

R-mMIMO \cite{MRA_CE} is a promising solution to enhance the MIMO capacity for a given antenna aperture. An R-mMIMO system can be obtained from a T-mMIMO system by replacing the conventional antennas with reconfigurable antennas. The basic idea of a reconfigurable antenna is to alter the physical structure of the antenna with the help of RF switches, so that the surface current density distribution becomes tunable. As an implementable solution based on state-of-the-art hardware fabrication methodologies, the EM properties of each R-mMIMO antenna can be made configurable within a limited set of operation modes. By contrast, although HMIMO has the theoretical capability to manipulate the whole continuous antenna aperture, it is physically unrealizable in practice.

\begin{figure}[!h]
\vspace*{-3mm}
\begin{center}
\includegraphics[width = 0.5\textwidth]{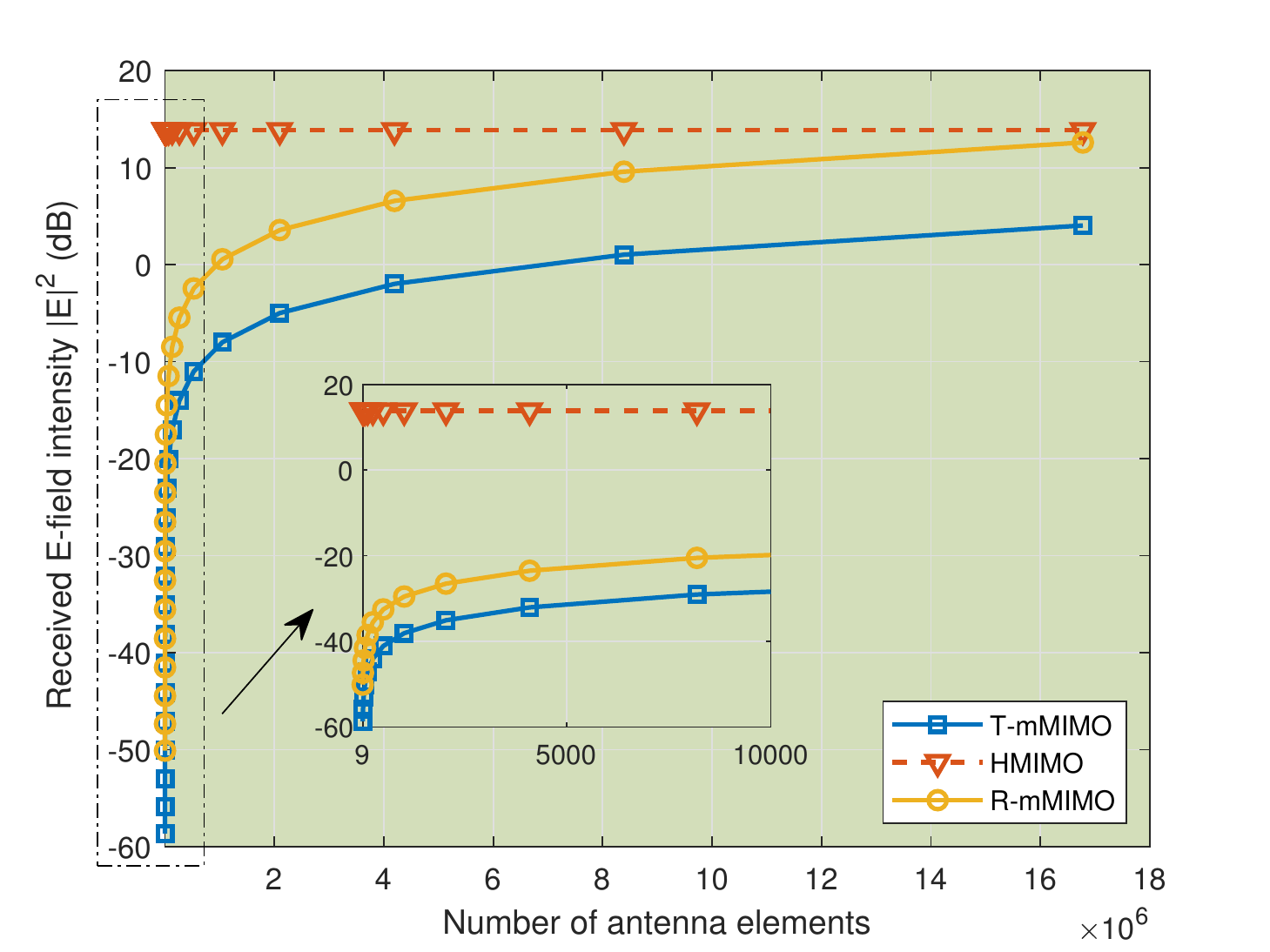}
\end{center}
\vspace*{-5mm}
\caption{Comparison of received E-field intensities for different MIMO systems. A fully-digital linear array with an aperture of 4$\lambda$ (operated at $3$ GHz with $\lambda$ denoting the wavelength) is placed along the y-axis with its geometric center at the origin. The target receiver is randomly located in the area $\left\{\left(x,y\right)|x\in  \left[5,50\right]\textrm{m}, y\in \left[50,100\right] \textrm{m}\right\}$, which is in the far-field region of the array. The results shown are averaged over 3,000 randomly generated target positions.}
\label{holo} 
\end{figure}

To provide an illustrative comparison of different types of MIMO systems, Fig. 1 considers a free-space propagation scenario, where a linear array is placed along the y-axis with its geometric center at the origin, and the received E-field intensity is simulated for the target receiver randomly located in a given far-field area. Optimal singular value decomposition-based precoding is performed at the transmitter. Without loss of generality, we use a Hertz dipole antenna to model each element of the T-mMIMO system for its tractable EMR expression, while the same antenna model with the additional capability to rotate the radiation pattern is assumed for R-mMIMO and HMIMO. As we assume a fixed array aperture, we reduce the antenna spacing when increasing the number of array elements. As can be observed, there is a performance gap between T-mMIMO and HMIMO even for very large numbers of antennas. This is because the radiation pattern shape of a Hertz dipole antenna is not omnidirectional, but a sinusoidal function in elevation direction. Observe that for large numbers of antennas, R-mMIMO can largely fill the performance gap between T-mMIMO and HMIMO by adjusting the main-lobe direction of each element's radiation pattern. For comparatively small numbers of antennas, which is of practical relevance, R-mMIMO can reduce the performance gap between T-mMIMO and HMIMO. With the capability of providing directional beam patterns to regions outside the normal direction, R-mMIMO can radiate more energy to the target position compared to T-mMIMO. In fact, the performance of ultra-dense R-mMIMO converges to that of HMIMO as the number of antenna elements goes to infinity. In this article, we are interested in the practical case of a limited number of antenna elements, where R-mMIMO can provide significant performance gains over T-mMIMO at an acceptable hardware cost.
 
The reconfigurability of the EM properties provides R-mMIMO with extra DoFs. In particular, in this article, we consider a reconfigurable pixel antenna (RPA) structure with radiation pattern reconfigurability. In T-mMIMO systems, the radiation pattern of each patch element is fixed, which means that the spatial coverage area of a pattern is always limited. By contrast, an RPA-based system can reconfigure the surface geometry of its parasitic layer for each antenna independently, so that the E-field can be manipulated to generate different radiation patterns. To integrate RPAs into a real-world wireless systems, this article propose to deploy arrays of RPAs, which leads to a new architecture for R-mMIMO. The detailed architecture of this proposed R-mMIMO is presented in the following section. This novel R-mMIMO is a candidate base station (BS) architecture providing enhanced precoding capabilities for future wireless communication systems. 

\begin{figure*}[!t]
\centering
\includegraphics[width = 0.9\textwidth]{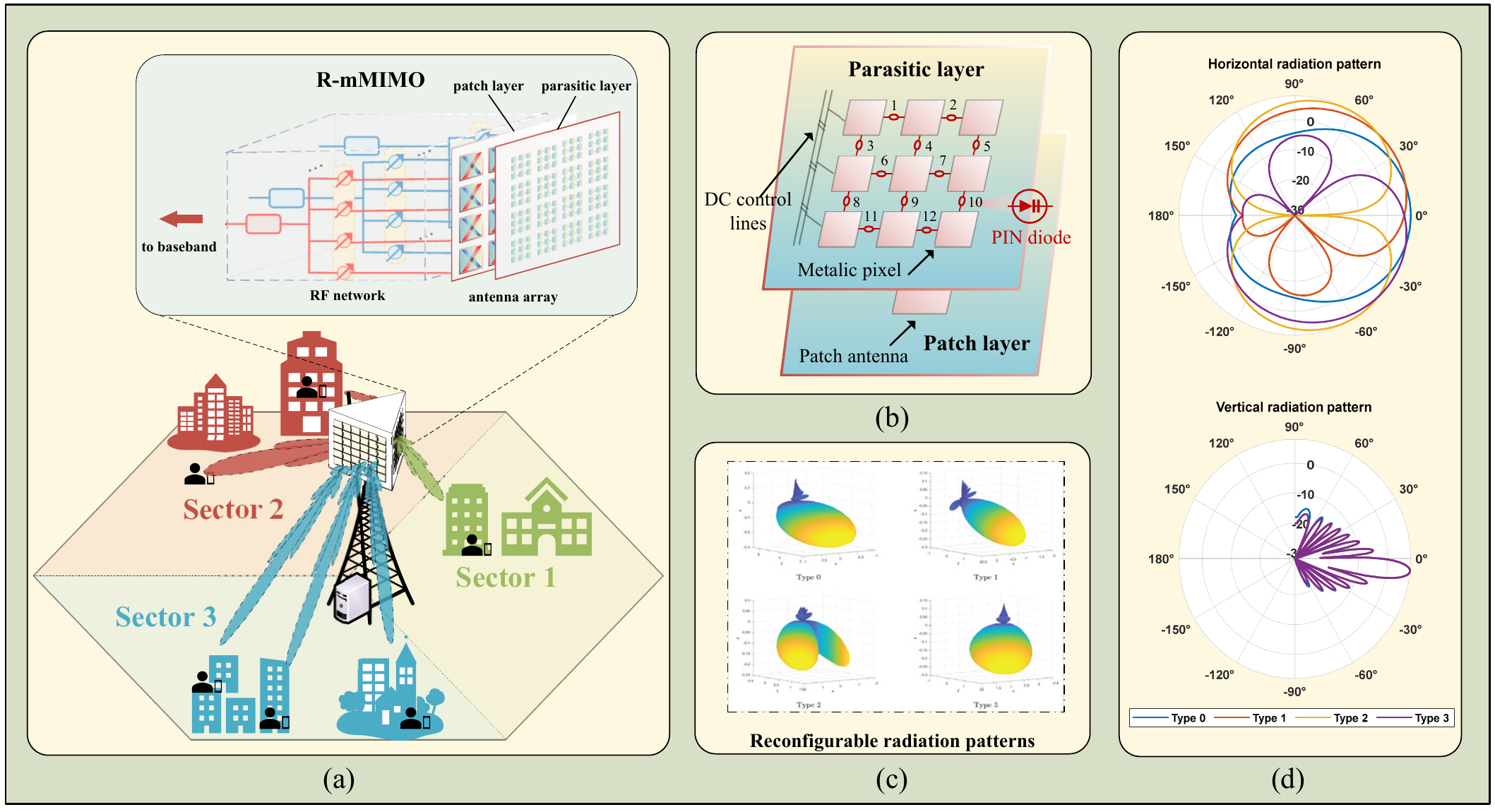}
\vspace*{-1mm}
\caption{Schematic diagram of R-mMIMO systems: (a)~multi-user downlink transmission and the corresponding SCA-based R-mMIMO architecture, (b)~structure of a single RPA, (c)~examples of 3D radiation pattern produced by an RPA, and (d)~2D horizontal and vertical radiation pattern cuts of~(c).}
\label{Array} 
\vspace*{-3mm}
\end{figure*}

\section{System Model of R-mMIMO}\label{S3}

\subsection{Architecture of R-mMIMO}\label{different} 

Fig.~\ref{Array}\,(a) illustrates the schematic diagram and the hardware design of an R-mMIMO system. In addition to the RF network and antenna array of T-mMIMO, R-mMIMO includes an extra parasitic layer on top of its patch layer. This additional parasitic layer allows each antenna to shape its own radiation pattern. Without loss of generality, we take SCA-based R-mMIMO as an example to illustrate the system modeling and evaluate performance, since it is easy to implement at low hardware cost for practical deployment and can be easily extended to the cases of FDA- and FCA-based R-mMIMO.

Fig.~\ref{Array}\,(b) depicts a single RPA, which consists of a patch layer and a parasitic layer. The patch layer carries the patch antenna, which couples the energy from the RF chain into space. The parasitic layer is composed of multiple inter-connected metallic pixels. The operating principle of such an RPA system can be described by the theory of reactively controlled directive arrays developed by Harrington \cite{Harrington}. This theory shows that the radiation pattern of the original antenna can be reconfigured through proper reactive loading of the parasitic elements. In the RPA system, the loading, i.e., the metallic pixels, are typically interconnected by electronically controllable switches, such as PIN diodes or RF micro-electro-mechanical-systems (RF-MEMS) \cite{Hardware2}. By setting the on-off status of the switches in the parasitic layer, the radiation pattern of each antenna can be flexibly reconfigured. 

The design of the parasitic layer is the key to the reconfigurability of the radiation pattern. An RPA offers limited adjustability of the current distribution on the surface of the parasitic layer, and the available patterns of each RPA are also constrained by the designed parasitic layer. In the example shown in Fig.~\ref{Array}\,(b), there are 12 PIN switches for the parasitic layer above each antenna, which can potentially produce $2^{12} = 4096$ EMR patterns. However, not all the patterns are suitable for information transfer. In \cite{GA}, the authors proposed an offline genetic algorithm to choose suitable PIN connections, so that the desired steering angles can be customized. This helps to remove the unsuitable patterns to only keep the desired patterns that can be used. The actual hardware design is beyond the scope of this article, and interested readers may refer to \cite{Hardware2} for more details.

Fig.~\ref{Array}\,(c) shows an example of the radiation patterns produced by a single RPA, and its 2D horizontal and vertical cuts are depicted in Fig.~\ref{Array}\,(d). Denoting the azimuth angle with respect to the local coordinate of the BS by $\phi$, we consider four types of patterns with different radiation directions:
\begin{itemize}
\item \textbf{Type 0}: Normal pattern with peak gain at $\phi=0^{\circ}$; 
\item	\textbf{Type 1}: Tilt pattern with peak gain at $\phi=30^{\circ}$; 
\item \textbf{Type 2}: Split pattern with peak gains at $\phi=\pm 56^{\circ}$;
\item \textbf{Type 3}: Tilt pattern with peak gain at $\phi=-30^{\circ}$. 
\end{itemize}
Intuitively, the EMR pattern of a T-mMIMO antenna has a fixed radiation direction like \textbf{Type 0}, which yields a lower coverage gain compared to R-mMIMO with multiple pattern choices. We note that in contrast to the simple Hertz dipole antenna considered in Fig. 1, practical RPAs can generate diverse radiation pattern shapes, such as the \textbf{Type 2} split pattern. Therefore, RPAs can also provide more diverse patterns than what is possible by rotating conventional antenna patterns.

In general, the extra parasitic layer does not incur much hardware cost and power consumption, since metallic pixels and electronically controllable switches are low-cost and consume low energy. Thus, due to the reconfigurability of its patterns, the considered SCA-based R-mMIMO is expected to achieve a comparable performance as FCA/FDA-based T-mMIMO but at a much lower hardware cost. We illustrate the performance of R-mMIMO for a typical urban macro (UMa) cell transmission scenario in the following subsection.

\subsection{Downlink Precoding with R-mMIMO}\label{S3.2}
We consider a downlink transmission system, where a BS with $N_t/2$ pairs of dual-polarized antennas and $M_t \le N_t$ RF chains simultaneously transmits $U\le M_t$ data streams to $U$ single-antenna user equipments (UEs). Without loss of generality, we assume that each UE employs an ideal omnidirectional antenna, and each BS array element is an RPA, whose radiation pattern can be configured within a pattern set of cardinality $P$. For example, the radiation patterns of Type 0~-~Type 3 in Fig.~\ref{Array}\,(c) constitute a reconfigurable pattern set with a cardinality of four. 

For serving the $U$ scheduled UEs, the baseband information signal vector is first precoded with an $M_t\times U$-dimensional digital precoder matrix $\bm{F}_{\mathrm{BB}}$. Next, the precoded vector is up-converted to the RF before passing through an analog precoder $\bm{F}_{\mathrm{RF}}$ of dimension $N_t\times M_t$. Finally, the RF signal is radiated into the channel which is affected by the tuning of the parasitic layer. To account for the radiation patterns of the antennas, let $\bm{h}_{u}\left(\bm{\mu}\right)$ denote the $N_t$-dimensional channel vector between the BS and the $u$-th UE, where the $N_t$-dimensional vector $\bm{\mu}$ indexes the adopted radiation patterns of the transmit RPAs. In other words, the effects of the EMR pattern are included in the channel vector, see also \cite{MRA_CE,QuaDRiGa}\footnote{A detailed description of the impact of the RPA on the channel cannot be provided here due to the limited space. Interested readers may refer to our online supplementary material for more details, see \url{https://github.com/kekeyingBIT/R-mMIMO/blob/main/supplement.pdf}}. 
Then, according to Shannon's formula, the overall SE of the system is given by 
	\begin{equation}\label{eq2}
		R\! = \sum\limits_{u=1}^{U}\! \log_{2}\! \left(\!\! 1\! +\! \frac{|\bm{h}_{u}^{ H}\left(\bm{\mu}\right)\bm{F}_{\mathrm{RF}}\bm{f}_{u}|^2}{\sum\limits_{j\neq u}^{U}\! |\bm{h}_{u}^{H}\left(\bm{\mu}\right)\bm{F}_{\mathrm{RF}}\bm{f}_{j}|^2\! +\! \sigma_n^2}\! \right)\!\! , \!
	\end{equation}
	where $\bm{f}_{u}$ is the digital precoder for the $u$-th UE, $\bm{F}_{\mathrm{BB}} = \left[\bm{f}_{1} ~ \bm{f}_{2} \cdots \bm{f}_{U}\right]$, and $\sigma_{n}^2$ is the power of the complex additive white Gaussian noise (AWGN) at the receiver.

In SCA-based T-mMIMO, hybrid precoding aims to optimize the analog and digital precoders for maximization of the SE. Typically, the non-zero entries of the analog precoder are constrained by the phase-shift network connections, i.e., they have constant modulus and finite phase resolution. Moreover, the Frobenius norm of the product of the analog precoder and the digital precoder is constrained by the transmit power. Numerous studies have been conducted to solve the resulting optimization problem so that the performance gap between SCA-based T-mMIMO and FDA-based T-mMIMO is minimized \cite{HBF_survey, Dynamic}. However, the SE optimization for R-mMIMO systems is complicated by the additional index vector $\bm{\mu}$, which we also refer to as EMR precoder in this article. The EMR precoder performs EMR precoding by selecting the radiation pattern for each RPA. Thus, a \textit{three-level precoding} involving the digital, analog, and EMR precoders is required. We note that SCA-based T-mMIMO can be considered as a special case of SCA-based R-mMIMO, where all the RPAs at the BS use the same legacy radiation pattern, for example, Type 0.

\section{Three-Level Precoding for R-mMIMO}\label{S4}

\subsection{Overview of Existing Schemes}\label{S4.1}

To optimize SE, the joint design of the digital, analog, and EMR precoders is needed. However, an efficient solution to this joint design problem is not yet available in the literature. The additional  discrete-valued EMR precoder complicates the problem, rendering the joint design a highly complex mixed continuous-discrete optimization problem, elaborated as follows: 1) the search space of the EMR precoder expands exponentially with the number of transmit antennas, and it is impossible to traverse all $P^{N_t}$ options due to the prohibitive complexity; 2) for each EMR precoder in the search space, the SE is a highly non-convex and non-differentiable function, and it is very difficult to find an approximate objective function to optimize at a reasonable computational cost while ensuring good performance.

Rather than tackling the intractable joint design, existing schemes usually focus on the optimization of the EMR precoder only, which we refer to as EMR domain precoding in this article. In \cite{MRA_BF}, the authors considered FDA-based R-mMIMO with multi-user transmission. By adopting zero-forcing fully-digital precoding, the received signal-to-noise ratio (SNR) at the UE can be equivalently used as an optimization metric to maximize the SE, and the authors of \cite{MRA_BF} proposed an iterative mode selection method to design the EMR domain precoding by maximizing this SNR metric. The work in \cite{TS} considered small-scale MIMO system with a single UE, where only EMR domain precoding was studied, and Thompson sampling and upper confidence bound algorithms were  applied for its design. Obviously, when extending to hybrid arrays, mMIMO, and multi-user scenarios, these EMR domain precoding methods cannot be directly applied. 

\begin{figure*}[!t]
\vspace*{-4mm}
\centering
	\subfigure[]{
		\includegraphics[width=0.333\textwidth, height= 0.2\textheight]{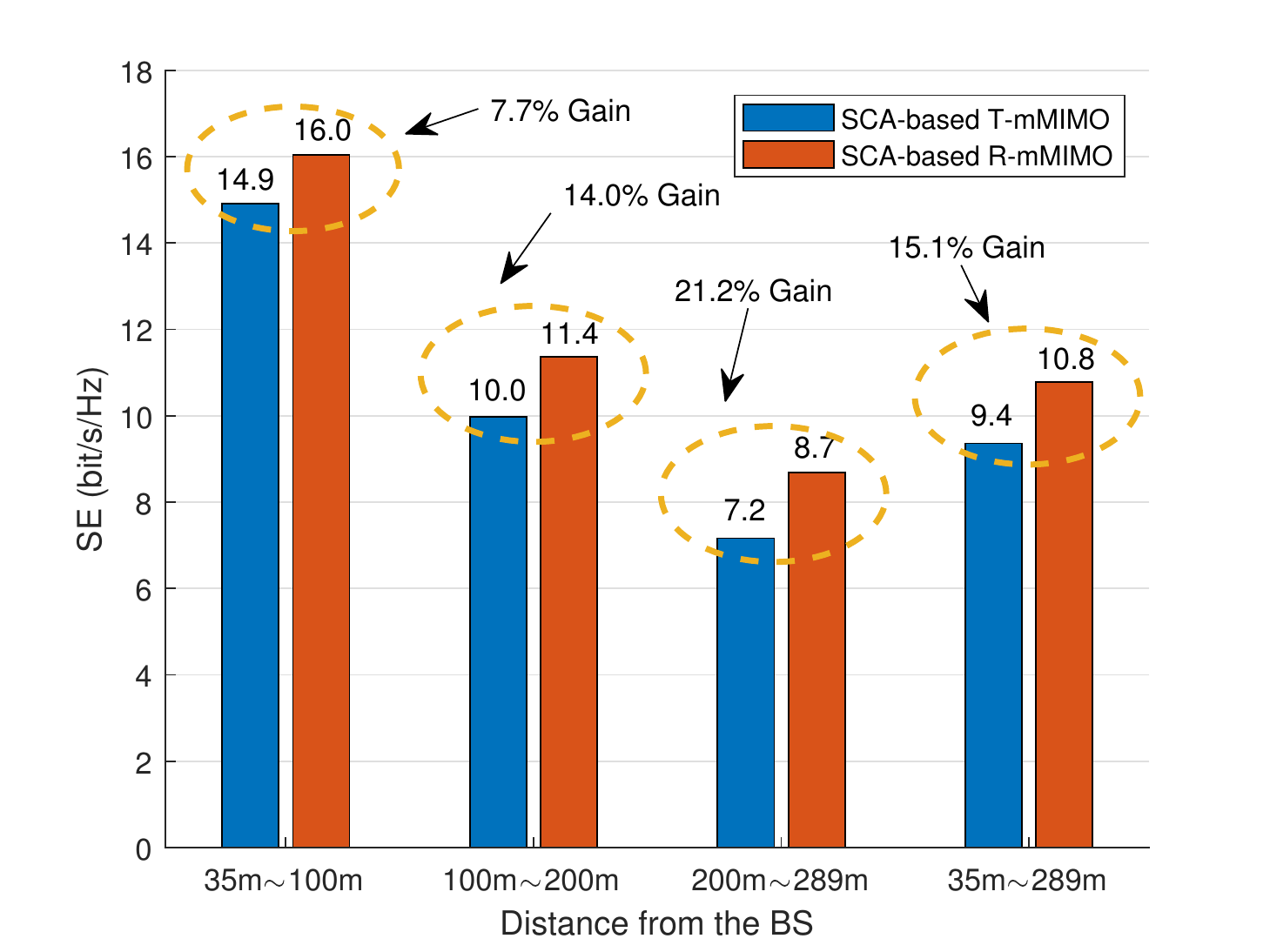}
	}
	\quad
	\hspace{-10mm}	
	\subfigure[]{
		\includegraphics[width=0.333\textwidth,height= 0.2\textheight]{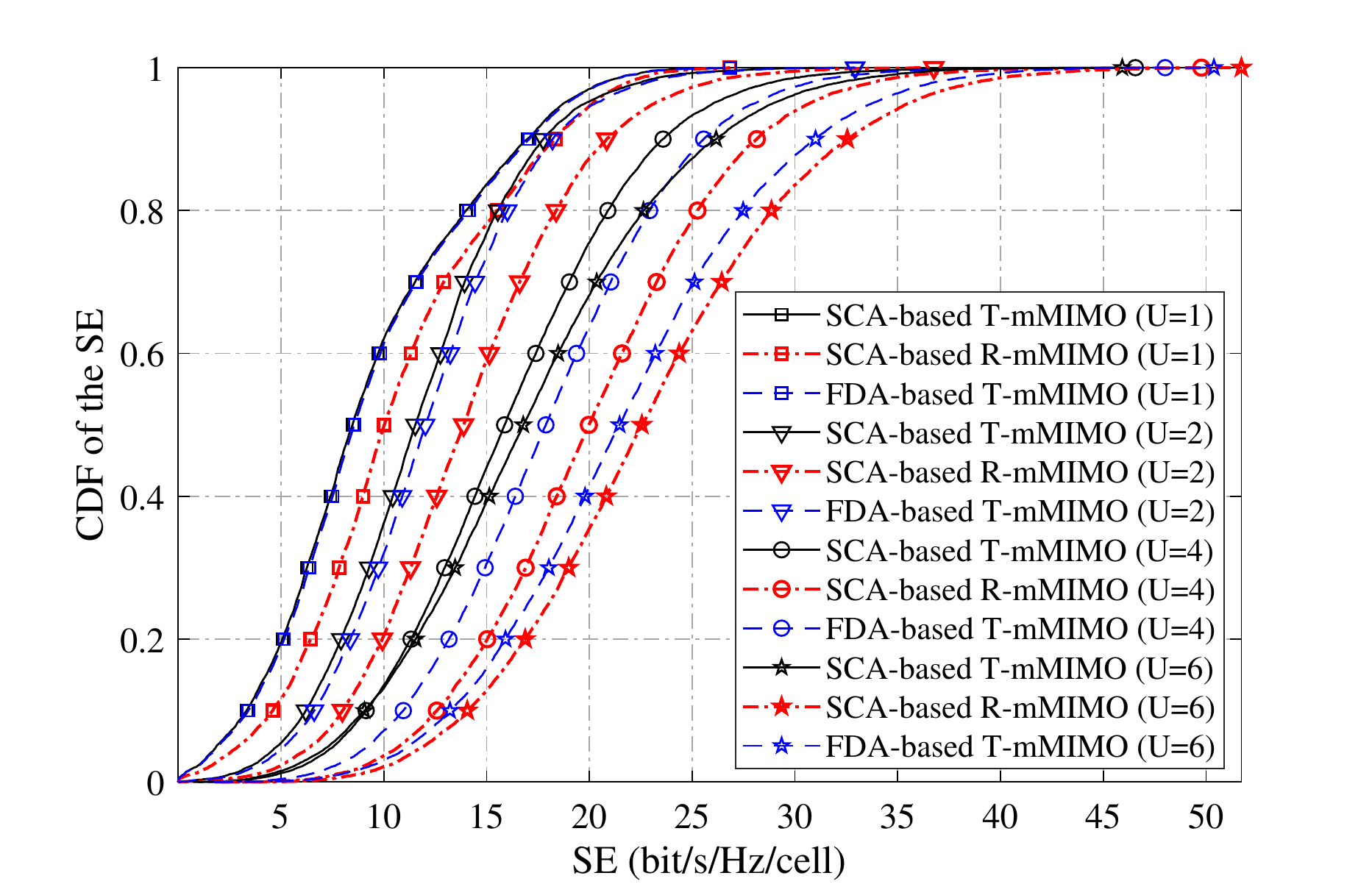}	
	}
	\quad
	\hspace{-10mm}
	\subfigure[]{
		\includegraphics[width=0.333\textwidth, height= 0.2\textheight]{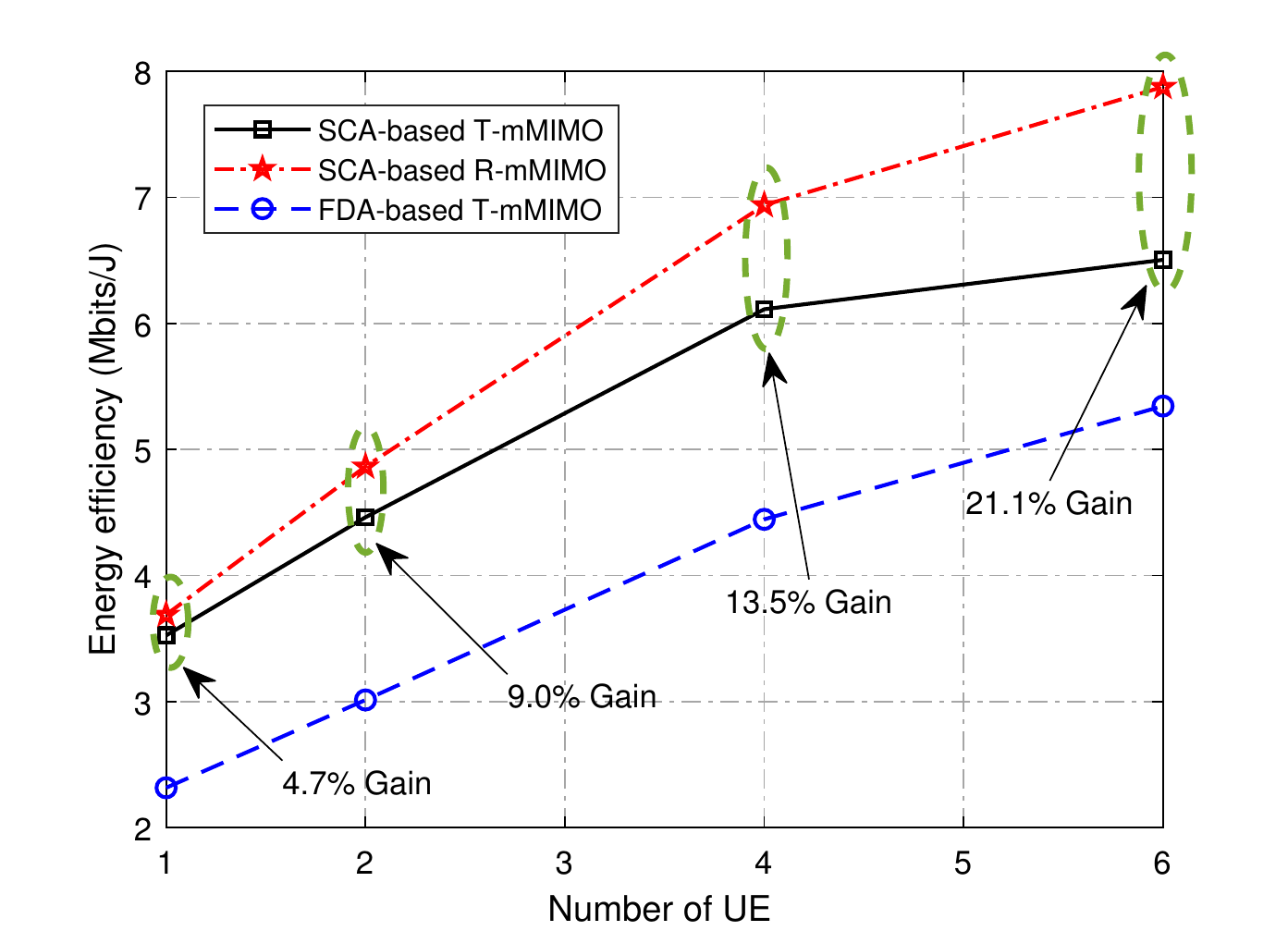}
	}
\vspace*{-2mm}
\caption{(a)~SE gains achieved by R-mMIMO for different geographic regions, (b)~CDF curves of the overall SE under different BS architectures, and (c)~EE of different BS architectures.}
\label{SEEE} 
\vspace*{-1mm}
\end{figure*}

\subsection{Three-Level SE Optimization}\label{S4.2}

To support multi-user and general wideband hybrid precoding in R-mMIMO systems, we further extend the iterative mode selection method of \cite{MRA_BF} to a generalized solution. Specifically, given the full channel state information (CSI) for all UEs, the proposed three-level SE optimization algorithm can be divided into two stages. In the first stage, an EMR domain precoding algorithm is employed to optimize the EMR precoder. Given the optimal EMR precoder obtained in the first stage, we jointly optimize the analog and digital precoders based on the resulting equivalent channel, as traditional hybrid precoding algorithms do. Next, we provide a greedy search algorithm for the first stage. The proposed greedy search algorithm for EMR domain precoding is summarized as follows. The search starts by initializing the pattern index of all transmit antennas to the legacy pattern, i.e., setting all the RPAs at the BS to Type 0. Then, in each iteration, the EMR patterns for each RPA are selected sequentially. To be more specific, in the $i$-th iteration, in order to optimize the $n_t$-th antenna's pattern, we keep all the other antennas' patterns fixed, and apply a standard hybrid precoding algorithm for all $P$ possible EMR patterns for the current antenna and evaluate the corresponding $P$ SE values. The pattern yielding the highest SE is selected for the $n_t$-th antenna. After all $N_t$ antennas have been configured one by one, the algorithm enters the next iteration. Typically, this greedy search method converges in $T_{iter} = 3$ to $5$ iterations.

Observe that with this EMR domain precoding algorithm, the number of searches for the EMR precoder is reduced to $N_t P T_{iter}$ compared to $P^{N_t}$ for the optimal exhaustive search. Therefore, a substantial amount of computational resources and time can be saved. However,  
in general, the resulting greedy-search based EMR domain precoding scheme only finds a good feasible solution for the EMR precoder rather than the optimal solution. We note that more sophisticated search strategies, such as evolutionary algorithms or swarm intelligence algorithms, can be applied to the constrained derivative-free optimization problem for the EMR precoder. These strategies can explore and exploit the search space more comprehensively. Nevertheless, our simulation results in Section. \ref{Sec.Sim} show that the proposed greedy algorithm already yields considerable performance gains over T-mMIMO in typical application scenarios. 

\section{Performance Comparison}\label{Sec.Sim}
In this section, we evaluate the performance of the proposed SCA-based R-mMIMO architecture and compare it with that of FDA- and SCA-based T-mMIMO architectures. We adopt the QuaDRiGa software package for channel generation \cite{QuaDRiGa}. As illustrated in Fig.~\ref{Array}\,(a), we consider downlink transmission in the UMa cell scenario. The BS employs an SCA with $N_t/2 = 16$ pairs of dual-polarized antennas and $M_t = 8$ RF chains.  
For each cell, a total of $15$ UEs are randomly distributed, and each UE is equipped with a single omnidirectional antenna. Penetration loss is taken into account for indoor UEs. For each transmission time interval (TTI), round robin scheduling is employed to select $U \le 8$ UEs for performing downlink precoding.  Furthermore, the overall transmit power for each cell is $42$\,dBm and the noise power spectral density is $-174$\,dBm/Hz. Throughout the simulation, inter-cell interference is neglected in order to keep our numerical comparison simple. For the baseline precoding algorithm, we adopt the eigen-zero-forcing method for FDA and the extended algorithm of \cite{Dynamic} for SCA, respectively. Moreover, a reconfigurable pattern set containing Type 0 to Type 3 from Fig.~\ref{Array}\,(d) is considered for SCA-based R-mMIMO ($P=4$), and a fixed pattern Type 0 is utilized for FDA/SCA-based T-mMIMO \footnote{Practical results \cite{MRA_BF} showed that RPAs achieve an additional antenna gain compared to conventional antennas. However, considering the introduction of an extra parasitic layer, we assume this additional antenna gain is fully compensated by the insertion loss. Therefore, we assume all radiation patterns have the same maximum radiation gain, as shown in Fig.~\ref{Array}~(d).}.  
  
The absolute SE gains achieved by SCA-based R-mMIMO over SCA-based T-mMIMO for UEs in different geographic regions are shown in Fig.~\ref{SEEE}\,(a). Note that $U = 1$ UE is served in each TTI for each cell. Here, we divide the geographic region into different parts according to the horizontal distance between the UE and the BS, and the average SE results for the near ($35$ m$\sim$$100$ m), middle ($100$ m$\sim$$200$ m), far ($200$ m$\sim$$289$ m) regions as well as the entire ($35$ m$\sim$$289$ m) region are presented as bar charts. As can be seen, the proposed R-mMIMO architecture achieves a higher SE than T-mMIMO for the UE at any distance. Intuitively, this can be explained by the fact that the EMR pattern used in this example is reconfigurable in the horizontal plane, and thus, a more flexible horizontal beam can be customized for a given channel environment.

Fig.~\ref{SEEE}\,(b) compares the cumulative distribution functions (CDFs) of the SE for the considered different array architectures, i.e., SCA-based T-mMIMO, FDA-based T-mMIMO, and SCA-based R-mMIMO. When the number of scheduled UEs is one, two, four, and six, respectively, the average SE gain of SCA-based R-mMIMO over SCA-based T-mMIMO is 15.1\%, 20.0\%, 24.9\%, and 33.3\%, respectively. Intriguingly, in this application scenario, the proposed scheme even achieves a better SE than FDA-based T-mMIMO. For any system architecture, the numbers of RF chains and antennas are fixed once the system design is determined. Therefore, the average SE performance for each UE decreases with the increasing number of scheduled UEs. The proposed R-mMIMO can provide extra precoding DoFs to mitigate this performance degradation to some extent. Consequently, the relative SE gain provided by R-mMIMO increases with the increased number of scheduled UEs, which indicates that the reconfigurability in the EMR domain can provide considerable SE gains. 

Next, we compare the EE performance of the three considered array architectures. The EE is defined as the SE divided by the total consumed power. The total consumed power includes the precoder power consumption and the transmit power consumption. For FDA-based T-mMIMO, the precoder power consumption is mainly caused by the $N_t$ RF chains, with each RF chain consisting of two digital-to-analog converters (DACs), two low-pass filters (LPFs), two mixers (MXs), and a local oscillator (LO) which is shared by all the chains. The power consumptions of these components are given by $P_{\mathrm{DAC}} = 200\,\mathrm{mW}$, $ P_\mathrm{{LPF}} = 14\,\mathrm{mW}$,  $P_\mathrm{{MX}} = 19\,\mathrm{mW}$, and $ P_\mathrm{{LO}} = 5\,\mathrm{mW}$, respectively \cite{EE_ref}. For SCA-based T-mMIMO, the precoder power consumption is mainly due to its $M_t$ RF chains and $N_t$ phase shifters. The power consumption of a single phase shifter (PS) is given by $P_\mathrm{{PS}} = 30\,\mathrm{mW}$ \cite{EE_ref}. In our proposed SCA-based R-mMIMO architecture, extra power is consumed by the parasitic layer. In this example, each RPA contains $12$ electrically-controlled switches with each switch (SW) consuming $P_\mathrm{SW} = 5\,\mathrm{mW}$  \cite{EE_ref}. Therefore, an additional power consumption of $60\,\mathrm{mW}$ is assumed for each RPA. The transmit power is set to $P_t = 14.4\,\mathrm{W}$, which is a typical value for the UMa cell scenario. It can be seen from Fig.~\ref{SEEE}\,(c) that the proposed SCA-based R-mMIMO architecture offers significant EE benefits over FDA-based T-mMIMO and SCA-based T-mMIMO. Also, as expected, FDA-based T-mMIMO has the lowest EE.

\section{Challenges and Future Directions}\label{benefits}

The proposed R-mMIMO architecture has the potential to revolutionize MIMO systems in future 6G networks. However, there exist some key challenges that must be addressed before large-scale deployment becomes feasible. 

\subsection{Theoretical Capacity}\label{S6.1}

R-mMIMO outperforms T-mMIMO owing to the additional DoFs in the EMR domain. Intuitively, these DoFs enable the transmitter to actively adjust the energy distribution between multiple paths to obtain an improved channel capacity. However, there is no available research that analyzes the fundamental capacity limits of R-mMIMO systems, and how to design optimal EMR patterns efficiently to approach this capacity is still unknown. 

As a trade-off between T-mMIMO and HMIMO, the design of R-mMIMO systems involves both signal processing and EMR pattern design, which calls for a fundamental analysis using tools from EM information theory \cite{EIT} for transmission modeling, DoF analysis, and performance evaluation. Pattern space analysis is expected to be one potential approach. In \cite{MRA_CE}, the authors adopted a Gram-Schmidt technique to decompose the radiation pattern into a set of orthogonal basic patterns, which allows the antenna radiation pattern to be decoupled from the rest of the channel environment. Based on this, the transmission model can be reduced to the multiplication of the equivalent channel and the signal in the pattern space. This facilitates tractable transmission modeling and fundamental theoretical analysis of generic R-mMIMO systems.

\subsection{Multifunctional Reconfigurable Antennas}\label{S6.2}

The RPA discussed in this article only considers the reconfigurability of the radiation pattern shape in the E-field. The authors in \cite{Hardware2} presented experimental results and design techniques for multifunctional reconfigurable antennas (MRAs), which offer a potential approach to design antenna hardware with the capability of providing other desirable EM properties. 

An MRA can integrate the reconfigurability of multiple domains (e.g., the frequency, pattern, and polarization domains) into a compact structure at a low cost. The EM behavior of MRAs can be changed by altering their physical or geometrical properties using some common tuning mechanisms, such as electronic devices like diodes and varactors, artificial metamaterials like microfluidics, liquid crystals, graphene, and so on. Through flexibly configuring their EM properties, MRAs can be used at both the BS and the UE to provide multi-domain diversity, co-channel interference mitigation, and reliable links with enhanced data rate in wireless communication systems.

\subsection{CSI Acquisition}\label{S6.3}

In R-mMIMO systems, CSI acquisition, which involves a completely new dimension owing to the reconfigurability of the EMR pattern, is a challenging task that must be resolved. It is impossible to estimate the full CSI at a time due to the fact that the training overhead increases linearly with the number of available EMR patterns. 

In \cite{MRA_CE}, the authors propose a combined channel estimation and prediction scheme, where only a subset of the EMR patterns are trained for estimation, and the untrained patterns are predicted exploiting the correlations between the different patterns. To further extend this idea, one can utilize more sophisticated approaches, such as compressed sensing or deep learning, to acquire the estimates of the channel parameters (such as angles, gains, and delays) from the legacy pattern. Then, the CSI for other reconfigurable patterns can be constructed using these estimated parameters. Another challenging issue is that a long training progress may cause outdated CSI, which can affect the accuracy of data detection and the efficiency of precoding. A potential solution is to take statistical CSI into account for more accurate channel estimation. Furthermore, in a time-varying environment, multi-user scheduling, CSI feedback, and channel tracking also need to be optimized accordingly, which deserves further study.

\subsection{Low Complexity Precoding}\label{S6.4}

This article considers a two-stage precoding algorithm for multi-user wideband transmission, which introduces additional complexity in the EMR domain precoding stage. Existing heuristic algorithms, such as the greedy search adopted in this article, may not meet practical complexity constraints since the SE objective function is computationally complicated and hard to approximate. Therefore, it is critical to investigate low-complexity EMR domain precoding algorithms without sacrificing too much of the SE performance. In addition to intelligent search algorithms, like evolutionary algorithms and swarm intelligence algorithms, one possible method is to model the EMR domain precoding as a decision process and use deep reinforcement learning methods. By interacting with the channel environment and learning from the data, the BS can learn a good radiation pattern selection strategy during exploitation and exploration. Moreover, neural networks can also be exploited to reduce the computation complexity of the objective function, which can help to simplify the algorithm design. Another possible solution is to model the explicit relationship between the radiation pattern and the channel. After optimizing the radiation pattern in a continuous space through derivative-based methods, the EMR precoder can then be acquired by quantizing the result to the nearest discrete grid value. Furthermore, interference from neighboring cells should not be overlooked when deploying R-mMIMO in wireless networks. Since the analog radiation patterns may cause interference to nearby cells when enhancing the coverage of the local cell, cooperative optimization of multi-cell precoding is required.  

\subsection{Integration with Other Technologies}\label{S6.5}
The extra DoFs of R-mMIMO allow us to transmit addtional information in the EMR domain or to customize more preferable channel conditions. Specifically, reconfigurable antennas have been exploited for mode shift keying transmission \cite{MSK}, which employed radiation patterns with low correlation, therefore achieving better detection performance than traditional spatial modulation. Some recent works proposed to use polarization domain DoFs for carrying information \cite{Polar}, where a polarization modulation scheme was developed to boost the system throughput. It is reasonable to expect that the joint exploitation of all DoFs offered by radiation patterns and polarization will be possible in the future given the rapid development of reconfigurable antennas. Moreover, R-mMIMO can be also applied in sensing and communication systems. For example, by customizing channels with lower cross correlations among potential targets, more accurate sensing can be achieved. Meanwhile, R-mMIMO can reshape the relative energy distribution among multipath components, which might be helpful in unfavorable propagation environments. Finally, R-mMIMO could be exploited to introduce randomness into the channel, which might be beneficial for covert communication based on channel randomization or radiation pattern hopping.

%

\section{Conclusions}\label{S7}
An innovative R-mMIMO system based on RPA was proposed in this article. A typical example of UMa downlink transmission was studied to demonstrate that SCA-based R-mMIMO with three-level precoding is capable of achieving SEs and EEs that are noticeably superior to those of existing SCA- and FDA-based T-mMIMO. In particular, our simulation results have confirmed that R-mMIMO can provide higher SE and EE gains with the increased number of scheduled UEs. Moreover, we have discussed critical challenges pertaining to R-mMIMO and presented new research directions towards making R-mMIMO a practical technology for 6G communication systems.

\end{document}